\input{aipcheck}

\documentclass[
 ,final            
  ]
  {aipproc}

\layoutstyle{6x9}

\newcommand{\pT}     {$p_{\rm{T}}$}
\newcommand{\pip}    {$\pi^{+}$}
\newcommand{\pim}    {$\pi^{-}$}
\newcommand{\pinull}    {$\pi^{0}$}
\newcommand{\xF}  {$x_{F}$}
\newcommand{\AN} {$A_{N}$}

\newcommand{\Deg} {$^{\rm{o}}$}

\begin{document}

\title{Single Spin Asymmetries in the BRAHMS Experiment}


\author{F.Videb{\ae}k for the BRAHMS collaboration}{
  address={Physics Department, Brookhaven National Laboratory}
}

\begin{abstract}
The BRAHMS experiment at RHIC has measured the transverse single spin asymmetries in
polarized pp induced pion production at RHIC. The results from the RHIC run-5 shows a significant
asymmetry for \pip~ and \pim~ at moderate \xF. The trend of the data is in agreement with lower energy results 
while the absolute values are surprisingly large. 
The \pT~ dependence is approximately  inversely propotional to \pT~ in agreement with the pQCD expectations.
\end{abstract}

\maketitle

In the last decade or so, measurements of transverse single spin asymmetries in pp collisions 
with polarized beams have attracted much theoretical and experimental interest. 
Results at low beam energies\cite{E704} ($\sqrt{s}$=20 GeV) show a sizeable asymmetry up to 
$30\%$ at relative large Feynman-$x$ (\xF) ~and at moderate \pT.
It was expected, naively, from  lowest order QCD estimates that the cross 
sections should have very little spin dependence. 
In order to get a non-zero asymmetry it is necessary to have a spin-flip amplitude, 
a phase difference in the intrinsic states,  and a non zero scattering angle. 
This makes it a higher order effect that can occur either in the initial state or in the final state parton scattering.

The asymmetry or analyzing power \AN~  is defined as  $(\sigma^{+}-\sigma^{-})/(\sigma^{+}+\sigma^{-})$, 
where $\sigma^{+(-)}$ is the spin dependent cross section for the scattering $pp\to\pi X$, 
and with the spin direction oriented up or down transversely to the scattering plane. 
The target is either un-polarized or the cross sections are averaged over target polarization states.
Experiment E704 at FNAL  \cite{E704} has shown that \break $\rm{A_{N}}$(\pip) $>$ $\rm{A_{N}}$(\pinull) $> 0 > $ $\rm{A_{N}}$(\pim).
In a short run in 2004 BRAHMS observed non-zero spin asymmetries 
for \pip~ and \pim~ at moderate \xF~\cite{BRAHMS_DIS05_pp}.
The STAR  experiment  observed a positive \AN~ for \pinull~ at large \xF~ at RHIC \cite{STAR_An_04}.

The BRAHMS experiment at RHIC is primarily designed and operated to make measurements 
of semi-inclusive spectra of identified hadrons over a wide range in rapidity and \pT.
The PID coverage for pions up to momenta of 35 GeV/$c$ and the option
to measure at $2.3$\Deg~  ($\eta \approx 4$) 
makes it well suited to study transverse Single Spin Asymmetries for identified pions at moderate \xF.
The present contribution presents the preliminary results of  \AN~ for \pip~ 
and \pim~ at moderate values of \xF~ in pp collisions at $\sqrt{s}$=200 GeV at RHIC from the higher statistics run-5.

The BRAHMS forward spectrometer consists of 4 dipole magnets, 5 tracking chambers, 
two Time-Of-Flight systems and a Ring Imaging Cherenkov Detector (RICH) for particle identification. 
The angular coverage of the spectrometer is from 2.3\Deg~ to 15\Deg,  and the solid angle is 0.8 msr.
Details of experimental setup can be found in \cite{BRAHMS_NIM}.
The acceptance of the BRAHMS spectrometer does not exactly correspond to a 
fixed angle. 
However, as shown in  Fig.~1 there still is a roughly linear relation between \pT~ and \xF~ 
 for  $\theta = 2.3\rm{^o}$ at the maximum field setting of 7.2 Tm in the spectrometer. 
Scattering angles of $2.3$\Deg~ and 4\Deg~ are shown on the figure as dashed lines.
Thus care should be taken when comparing to both other experiments and to theory.

\begin{figure}[ht]
  \includegraphics[height=.20\textheight]{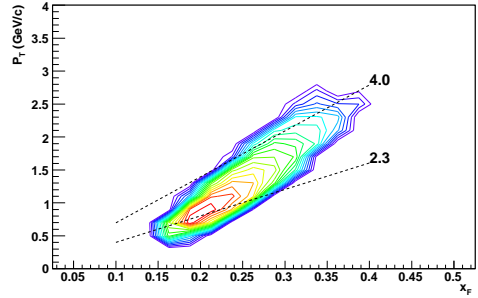}
  \caption{Acceptance in the BRAHMS experiment for pions at the 
nominal setting of 2.3\Deg in \pT~ vs.  ~\xF. The dashed lines indicates the \pT -  \xF~  
correlations for fixed angles of 2.3\Deg~ and 4\Deg, respectively.}

\end{figure}

Complete tracks were reconstructed from local track segments in at least 4 of the 5 chambers. 
The momentum of the track was obtained from 3 independent measurements. 
The momentum resolution $\delta p/p$ is estimated to be $\approx 1\%$ at momenta of 22 GeV/c. 
The tracks are required to project cleanly through the spectrometer.
An approximate vertex can be determined from the timing measurements in sets of symmetrically placed
Cherenkov counters (CC) around the beam pipe at 1.9 and 6.4 meters from the nominal interaction point.
The position resolution of the vertex determination is about 2 cm. 
In addition  live rates for individual bunches are obtained from these counters, from another set of Cherenkov 
Counters (BB) with limited acceptance at $\pm 2.15$m,
and from a pair of Zero Degree Calorimetres (ZDC) placed at $\pm 18$m.  
The tracks accepted in the spectrometer are required to point backward to these measurements with an accuracy of 15 cm and
to be within a narrow range of (-40,20) cm of the nominal interaction point.
Due to the measuring angle of 2.3\Deg~  the spectrometer tends to accepts track weighted 
towards negative vertex positions. 
The particle identification of  pions is done exclusively using the RICH. 
It is required that the
calculated radius for a pion  is within .25 cm from the measured radius, 
and at the same time more than .30 cm away from the estimated radius for a kaon.
This corresponds to about a 2 and 2.5 $\sigma$ cuts, respectively.

In the RHIC accelerator the transverse spin direction  is altered between the
bunches of polarized protons that forms the beam in each of the two rings. 
Thus most experimental time-dependent effects originating from the spectrometer, beam variations, and the
vertex determination cancel out when constructing the raw asymmetries
 $$\epsilon =(N^{+}-L*N^{-})/(N^{+}+L*N^{-})$$ The $N^{+(-)}$ represents the yield of pions in a given 
kinematic bin where the beam spin direction is up or down relative to the scattering plane determined by
$\vec{k}_{beam}\times \vec{k}_{out}$. 
The factor $L$ is the ratio of the luminosity of bunches with positive polarization to 
those of negative polarization thus accounting for  
 non-uniform bunch intensities.
The luminosity ratio is determined independently from the spectrometer data using several
measures of collision rates from the CC, BB, and ZDC detector systems. 
It is estimated that the systematic error from the relative luminosity  measurements is in order of $0.5\%$.
\begin{figure}[h]
 \begin{minipage}[t]{60mm}
\centering
  \includegraphics[width=6.5cm]{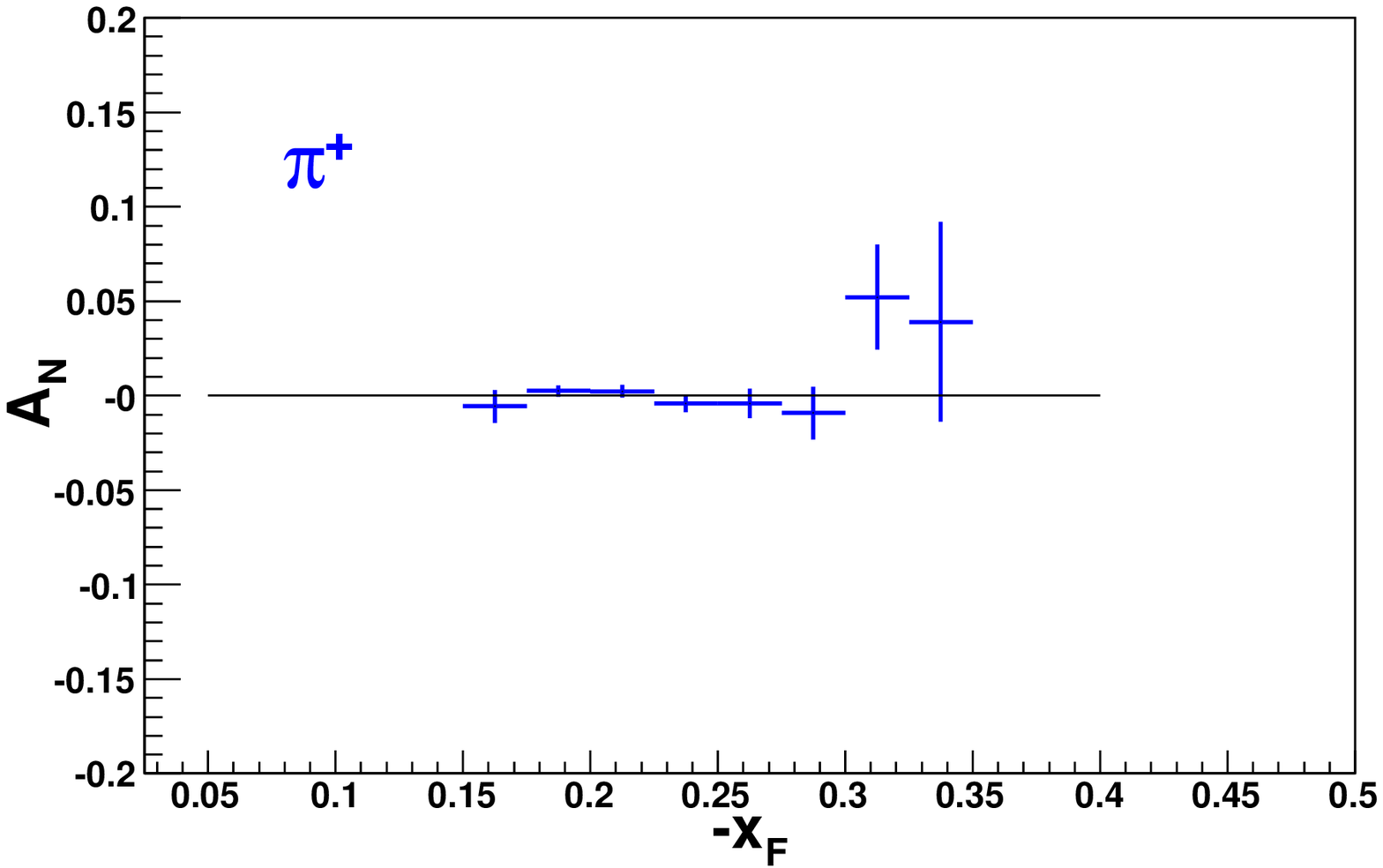}
  \caption{Analyzing power \AN~  \pim~ at negative \xF.}
\end{minipage}
\hspace{\fill}
 \begin{minipage}[t]{60mm}
\centering
  \includegraphics[width=6.5cm]{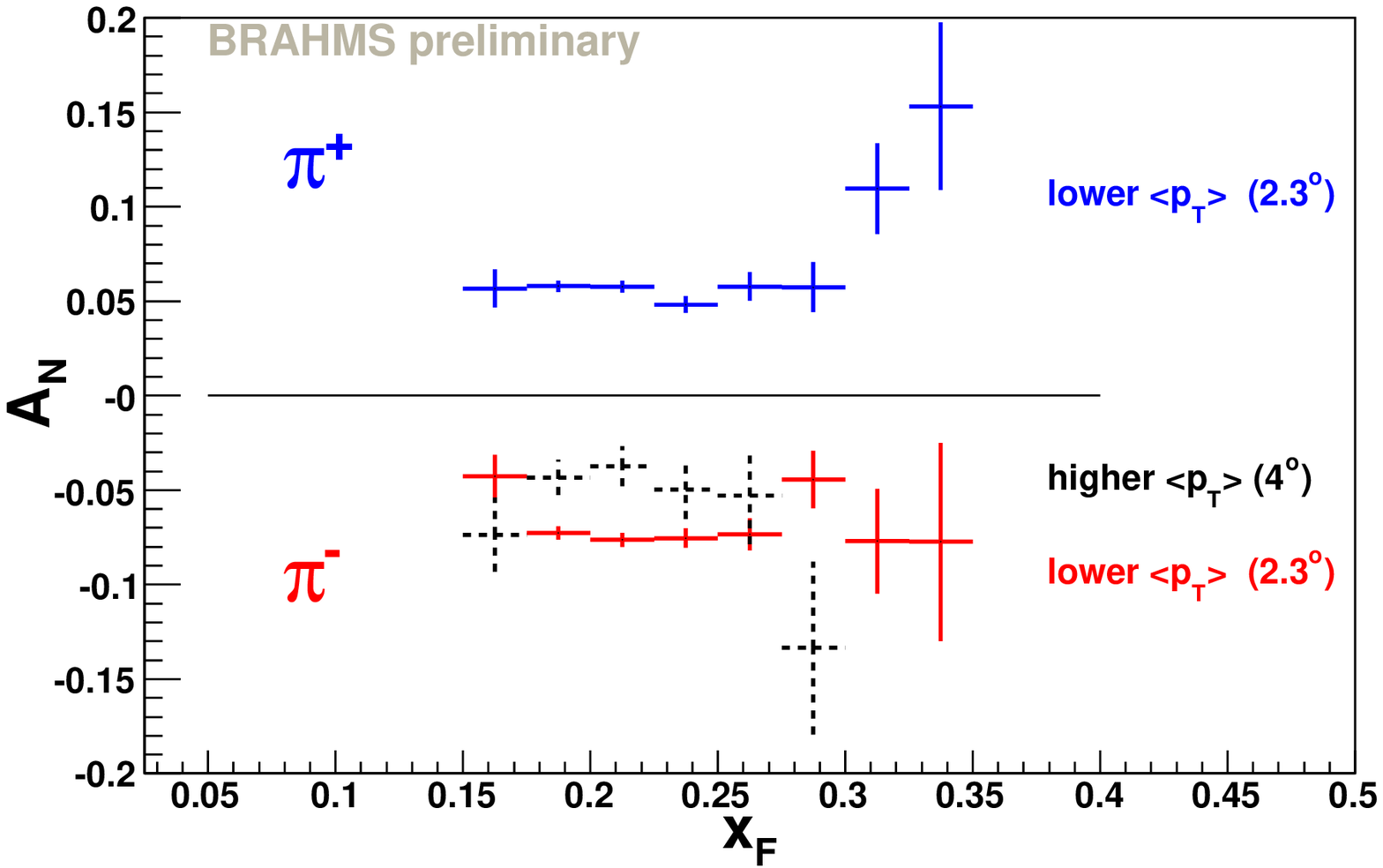}
  \caption{Analyzing power \AN~ for \pip~ and \pim~ at positive \xF~ (right hand panel) and negative \xF~ (left panel).
See text for details.}
\end{minipage}

\end{figure}

The asymmetry is then in turn determined from \AN$=\epsilon/P$. 
The polarization ($P$) as determined from the online CNI measurements \cite{CNI} is $\approx 50\%$ for 
the RHIC stores used in the data analysis 
The systematic error on beam polarizations is $\approx  20\%$ and represents a scaling error on the values of
\AN. 
This error is expected to be reduced after the final analysis of CNI data. 
The measured  asymmetries corrected for the beam polarization are shown in  Fig.~2 for \pim~ and \pip~ vs. \xF. 
The data for \pim~ in the right panel were obtained from measurement at 2.3\Deg (solid lines) and 4\Deg~ (dashed) , respectively. 
The 4\Deg~ data corresponds at the same \xF~ to a higher \pT~ value. 
In the \xF~ range of of 0.15 to 0.3 the mean \pT~ increases by approximately 40\% from 1.4 to 2.2 GeV/c.
In the same panel are shown the results for \pim~ for the 2.3\Deg~ data with values that are of opposite sign 
and approximately the same absolute value as \pim~ ( possibly slightly smaller).
By utilizing the polarization of the beam in the yellow ring  moving away from the spectrometer one can determine the asymmetry for \pip at
negative \xF~ values.
The resulting asymmetries are shown in the left panel of  Fig.~2 demonstrating that near zero values are obtained for \pip~ in
the range $-0.35<$\xF$<-0.15$. The results are consistent with that obtained for \pinull~ by STAR\cite{STAR_An_04}.

In summary, in the RHIC run-5 BRAHMS has measured  \AN~ 
for \pip~ and \pim~ with signs as observed previously in E704 at FNAL. 
Comparing data from two angle settings we show that
 the magnitude of the asymmetry falls with \pT~ approximately proportional to $1/ p_T$.

This work is supported by the Division of Nuclear Physics of the Office of Science of 
the U.S. Department of energy under contract DE-AC02-98-CH10886.

\end{document}